\documentstyle[12pt]{article}

\newcommand{\Om}{\Omega}

\newcommand{\al}{\alpha}
\newcommand{\ep}{\epsilon}

\newcommand{\la}{\lambda}

\newcommand{\deebar}{\bar{\partial}}

\newcommand{\msc}[1]{\mbox{\scriptsize #1}}
\newcommand{\dsp}{\displaystyle}

\newcommand{\br}{\mbox{{\bf R}}}
\newcommand{\bz}{\mbox{{\bf Z}}}

\newcommand{\cO}{{\cal O}}
\newcommand{\cN}{{\cal N}}

\newcommand{\ket}[1]{{|#1\rangle}}
\newcommand{\bra}[1]{{\langle#1|}}

\newcommand{\ktp}{\ket{0;+}}
\newcommand{\ktm}{\ket{0;-}}
\newcommand{\ktpm}{\ket{0;\pm}}
\newcommand{\btp}{\bra{0;+}}
\newcommand{\btm}{\bra{0;-}}
\newcommand{\btpm}{\bra{0;\pm}}

\newcommand{\mod}{\mbox{mod}}

\newcommand{\p}{\mbox{p}}
\newcommand{\ps}{\msc{p}}
\newcommand{\x}{\mbox{X}}
\newcommand{\xs}{\msc{X}}

\newcommand{\tG}{\tilde{G}}
\newcommand{\tL}{\tilde{L}}

\newcommand {\eqn}[1]{(\ref{#1})}

\makeatletter
\@addtoreset{equation}{section}

\makeatother

\begin{document}

\vskip 7mm
%%% Title page %%%%%
\begin{titlepage}
 
 \renewcommand{\thefootnote}{\fnsymbol{footnote}}
 \font\csc=cmcsc10 scaled\magstep1
 {\baselineskip=14pt
 \rightline{
 \vbox{\hbox{hep-th/0011148}
       \hbox{UT-917}
       }}}

 \vfill
 \baselineskip=20pt
 \begin{center}
 \centerline{\Huge  D-branes in} 
 \vskip 5mm 
 \centerline{\Huge Singular Calabi-Yau $n$-fold}
  \vskip 5mm 
 \centerline{\Huge and $N=2$ Liouville Theory}

 \vskip 2.0 truecm

\noindent{\it \large Tohru Eguchi and Yuji Sugawara} \\
{\sf eguchi@hep-th.phys.s.u-tokyo.ac.jp~,~
sugawara@hep-th.phys.s.u-tokyo.ac.jp}
\bigskip

 \vskip .6 truecm
 {\baselineskip=15pt
 {\it Department of Physics,  Faculty of Science\\
  University of Tokyo\\
  Bunkyo-ku, Hongo 7-3-1, Tokyo 113-0033, Japan}
 }
 \vskip .4 truecm

 \end{center}

 \vfill
 \vskip 0.5 truecm

\begin{abstract}
\baselineskip 6.7mm

Making use of the $\cN=2$ Liouville theory and world-sheet techniques, 
we study the properties of D-branes 
wrapped around vanishing SUSY cycles of singular Calabi-Yau 
$n$-folds ($n=2,3,4$). After constructing boundary
states describing the wrapped branes, we evaluate 
the disc amplitudes corresponding to the periods of SUSY cycles. 
We use the old technique of KPZ scaling in 
Liouville theory and derive holomorphicity
and scaling behavior of vanishing cycles which are in agreement with
geometrical considerations. 

We also discuss the open string Witten index using
the $\cN=2$ Liouville theory and obtain the intersection
numbers among SUSY cycles which also 
agree with geometrical expectation.

\end{abstract}

\setcounter{footnote}{0}
\renewcommand{\thefootnote}{\arabic{footnote}}
\end{titlepage}

\newpage

\section{Introduction}

Recently the study of string compactification 
on singular Calabi-Yau manifolds has been 
receiving a lot of attentions \cite{Strominger,GV,OV,GVW,GKP,SV}. 
When a CY manifold
approaches a singular limit, some of its cycles become degenerate
and various non-perturbative phenomena take place.
In particular when CY manifold $X_n$ ($n$ denotes the complex 
dimension) has an isolated A-D-E type 
singularity, we expect to obtain a scale invariant theory in $d=10-2n$ 
dimensions which is decoupled from gravity \cite{GVW}. 
It has been proposed that such a 
space-time theory may be studied by looking at its dual string theory 
\cite{ABKS,GKP,GK,Pelc}: this dual theory possesses a linear dilaton
background and involves the ${\cal N}=2$ Liouville theory \cite{KutS}
(or $SL(2,{\bf R})/U(1)$
theory) and the ${\cal N}=2$ minimal model.
Liouville field $\phi$ describes the radial direction of the
throat of singular CY manifold
and another scalar field $Y$ in ${\cal N}=2$ Livouville sector 
parameterizes the circular direction of the throat. 

Such a dual string theory is reminiscent of the Gepner model of
exactly soluble string compactification \cite{Gepner}. 
In the present case the minimal model is coupled to Liouville sector 
which is non-compact
and the non-compactness makes it non-trivial to see if such a string background
is in fact consistent. Recently, modular invariance for these backgrounds 
has been studied and it has been shown that there exist a series 
of modular invariant partition functions of the A-D-E type 
which are in one to one correspondence with the
CY manifolds with A-D-E singularities
\cite{ES,Mizoguchi,Yamaguchi,NN}.

It is quite interesting that the Liouville theory plays 
a novel role in the description of singular CY compactification. 
In this paper we will borrow the technique from the old Liouville theory 
\cite{2dgrav,2dgrav-review,Liouville} 
and apply it to the analysis of the scaling 
behavior of singular CY manifold.
In particular we will study the scaling laws of the periods
of vanishing cycles as the deformation parameter $\mu$ of the singularity
is turned off. 

In order to carry out such an analysis we use the method of boundary 
states which describes the SUSY cycles from the world-sheet point of view. 
Technologies of describing D-branes 
using world-sheet theory have recently been developed considerably 
\cite{OOY,BCFT}. In particular the boundary state approach to 
D-branes in Gepner models has been discussed by various authors
\cite{BCFT-Gepner,BDLR,BCFT-CY}. 
In this paper we will make a slight generalization of this construction 
to cope with the case of non-compact CFT consisting of Liouville field. 

We will consider two-point functions of boundary states with chiral 
primary fields which correspond to periods of vanishing cycles. 
These correlation functions vanish identically due to Liouville momentum
conservation unless the system is perturbed by some Liouville potential
terms. We shall show that when the theory is perturbed by the cosmological 
constant operator, correlation functions become non-zero and behave
exactly as the periods of vanishing cycles.
We shall 
show that we can in fact identify the deformation parameter $\mu$ as the 
coefficient of the cosmological constant term in Liouville theory.
We also compute the intersection numbers among 
vanishing cycles using the Liouville
theory. We obtain results which agree completely with geometrical 
considerations and also agree with the results of \cite{Lerche,LLS}
based on the $SL(2,{\bf R})/U(1)$ approach.

This paper is organized as follows: In section 2, starting from 
a brief review of the setup of world-sheet description of \cite{GKP}, 
we  construct the supersymmetric boundary states.
In section 3 we will evaluate the periods of the D-branes wrapped
around the vanishing cycles by means of the world-sheet techniques.
In section 4 we compute the intersection matrix (open string Witten
index \cite{DFiol}) of vanishing cycles. 
We will also discuss the relation of Liouville theory 
to the $SL(2;\br)/U(1)$ approach in section 5.
In section 6 we summarize our results 
and present some discussions.

\section{World-sheet Description of D-branes in 
Singular CY Compactification}

\subsection{World-sheet Approach to Singular CY Compactification}

Throughout this paper we consider type II string theory on the background
$\br^{d-1,1}\times X_n$, where $X_n$ is a CY $n$-fold ($2n+d=10$)
with an isolated singularity of A-D-E type defined by a polynomial equation 
$F(z_1,z_2,\ldots, z_{n+1})=0$. (More precisely, we shall consider
IIA string for the cases $d=2,~4$, and IIB string for $d=4$).
For simplicity we focus on the $A_{N-1}$-type singularity in this paper;
\begin{equation}
F(z_1,\ldots, z_n) = z_1^N + z_2^2+\cdots+ z_{n+1}^2 .
\end{equation} 
According to \cite{GKP} the decoupled limit of string theory 
on $\br^{d-1,1}\times X_n$ is described as a string theory on
$$
\br^{d-1,1} \times (\br_{\phi}\times S^1) \times
LG(W=F)  
\cong \br^{d-1,1} \times (\br_{\phi}\times S^1) \times M_N
$$
where $LG(W=F)$ stands for the ${\cal N}=2$ Landau-Ginzburg model with a 
superpotential $W=F$, and $M_N$ denotes the ${\cal N}=2$ minimal 
model with central charge $c=3(N-2)/N$.
The sector $\br_{\phi}\times S^1$ is described by 
the ${\cal N}=2$ Liouville theory \cite{KutS}
whose field contents consist of bosonic variables 
$\phi$ (parameterizing $\br_{\phi}$), $Y$ (parameterizing  $S^1$)
and their fermionic partners $\Psi^+$, $\Psi^-$. 
"$\br_{\phi}$" indicates a linear dilaton background
with the background charge $Q (\,>0)$\footnote
   {The present normalization of the background charge 
  is defined so that the Liouville central charge equals $c_{\phi}=1+3Q^2$.}.
Adjustment of the central charge gives the following
values of the background charge in dimensions $d=6,4,2$
\begin{eqnarray}
&&  d=6;\hskip3mm   \dsp Q=\sqrt{\frac{2}{N}}, \nonumber \\
&&  d=4;\hskip3mm  \dsp Q=\sqrt{\frac{N+2}{N}}, \\
&&  d=2; \hskip3mm \dsp Q=\sqrt{\frac{2(N+1)}{N}}.\nonumber 
\label{Q value}
\end{eqnarray} 
The superconformal
currents read as 
\begin{equation}
\left\{
\begin{array}{l}
\dsp  T =-\frac{1}{2}(\partial Y)^2 
-\frac{1}{2}(\partial \phi)^2 -\frac{Q}{2}\partial^2\phi
-\frac{1}{2}(\Psi^+\partial \Psi^- -\partial \Psi^+ \Psi^-), \\
\dsp  G^{\pm}=-\frac{1}{\sqrt{2}}\Psi^{\pm}(i\partial Y \pm \partial \phi )
\mp \frac{Q}{\sqrt{2}}\partial \Psi^{\pm}, \\
\dsp J= \Psi^+\Psi^- -Qi\partial Y ,
\end{array}
\right.
\label{SCA-L}
\end{equation}
which generate the ${\cal N}=2$ superconformal algebra (SCA) with 
$\dsp \hat{c}(\equiv\frac{c}{3})=1+Q^2$.

Since we have a linear dilaton background, $\dsp \Phi(\phi) = -
\frac{Q}{2}\phi$, the theory is weakly coupled in 
the "near boundary region" $\phi\sim +\infty$.
On the other hand, in the opposite end $\phi \sim -\infty$ 
(near the singularity) the string coupling blows up.
In order to avoid the region $\phi \sim -\infty$ 
one should add a "Liouville potential" term 
$\sim e^{-\alpha \phi}$ ($\alpha >0$)
to the action. Compatibility with 
the $\cN=2$ superconformal symmetry leads us to the following three types 
of Liouville potentials, two of them are (anti-)chiral 
and one is non-chiral;
\begin{eqnarray}
S^{\pm}&=&  \int d^2z\, \Psi^{\mp}\tilde{\Psi}^{\mp}
e^{-\frac{1}{Q}(\phi\pm i Y)},  \label{Liouville potential}\\
S_{\msc{nc}} &=& \int\,d^2z\, (\partial \phi-i\partial Y-Q\Psi^+\Psi^-)
 (\deebar \phi+i\deebar Y+Q\tilde{\Psi}^+\tilde{\Psi}^-)\,e^{-Q\phi} .
\label{Liouville potential 2}
\end{eqnarray}
They commute with all the generators of $\cN=2$ superconformal algebra
and thus are the screening charges of SCA. 
It is easy to recognize these as the super- and K\"{a}hler potential terms
in the ${\cal N}=2$ theory 
\begin{eqnarray}
S^{\pm} &=& \int d^2z d^2\theta^{\pm} \, e^{-\frac{1}{Q}\Phi^{\pm}}, \\
S_{\msc{nc}} &=& \int d^2z d^2\theta^+d^2\theta^-\,
    e^{-\frac{Q}{2}(\Phi^+ + \Phi^-)} , 
\end{eqnarray}
where $\Phi^{\pm}$ denotes the (anti-)chiral superfield whose scalar
component is $\phi\pm iY$. 
Thus the operators $S^{\pm}$ describe the deformation of the superpotential 
or the complex structure of
the CY manifold $X_n$ while
$S_{\msc{nc}}$ provides 
the deformation of the K\"{a}hler
structure of the theory.  
We can identify the operators 
$S^{\pm}$ as the analogues of the cosmological constant term in 
the bosonic Liouville theory. On the other hand, as we discuss later,  
the operator $S_{\msc{nc}}$
may be identified as the screening operator of $SL(2,{\bf R})$ current algebra 
or the black hole mass operator 
in the $SL(2,{\bf R})/U(1)$ WZW model of 2D black hole \cite{2DBH}.

In the following we shall concentrate on the perturbation
by the cosmological constant operators $S_{\pm}$ ($S_L$ denotes the Liouville
action)
\begin{equation}
S_L \Longrightarrow S_L +\mu S^+  + \bar{\mu} S^-.
\label{Liouville def}
\end{equation}
Note that $\mu$ here is a complex 
parameter as opposed to the bosonic Liouville theory where 
$\mu$ is a real parameter.

It has been suggested by various authors \cite{GV,GKP} 
that the perturbation (\ref{Liouville def}) 
corresponds to  
the most relevant deformation of the singularity
\begin{equation}
F(z_1, \ldots, z_{n+1} )+ \mu=0 .
\end{equation}
If we consider the case of $n=3$ for instance, 
superysmmetric 3-cycles in CY threefold are degenerate 
in the singular limit $\mu=0$. 
These cycles become inflated and acquire a finite size when the perturbation 
$\mu$ is turned on. 
In the following sections we compute correlation functions
in Liouville theory  
which correspond to the sizes and intersection numbers
of the vanishing cycles. These correlation functions in fact vanish 
by the Liouville momentum conservation at $\mu=0$, however, 
they acquire finite values
when the system is perturbed by the cosmological constant operator.
We use the technique of the Liouville theory developed in 2D gravity theory
\cite{2dgrav,Liouville} and  
compute the correlation functions by inserting a suitable number of 
cosmological constant operators.
It turns out that we obtain scaling relations for the vanishing cycles 
which are in complete agreement with the geometrical 
considerations.

%\footnote
%      {We are here taking the convention of screening charges $S^{\pm}$
%        different from \cite{ES}.}

\subsection{Supersymmetric Boundary States in $N =2$ Liouville Theory}

It is known that in the ${\cal N}=2$ theory
there are two types of boundary
states preserving the superconformal symmetry \cite{OOY}.
In the context of superstring theory
they correspond to the D-branes wrapped around 
special Lagrangian submanifolds or holomorphic cycles
in the CY manifolds. 
We describe them using the mode expansion in the closed string channel.

\begin{description}
 \item[A-type] (middle-dimensional cycles, special Lagrangian submanifolds)
\begin{eqnarray}
  (J_n-\tilde{J}_{-n}) \ket{B;\eta} &=& 0 \label{JA} \\
  (G^{\pm}_{r}-i\eta \tilde{G}_{-r}^{\mp})\ket{B;\eta} &=& 0 \label{GA} 
\end{eqnarray}

 \item[B-type] (holomorphic even-dimensional cycles)
\begin{eqnarray}
  (J_n+\tilde{J}_{-n}) \ket{B;\eta} &=& 0  \label{JB}\\
  (G^{\pm}_r-i\eta \tilde{G}_{-r}^{\pm})\ket{B;\eta} &=& 0  \label{GB}
\end{eqnarray} 
\end{description}
where $\eta= +1$ or $-1$. Signature $\eta$ is related to 
the choice of NS or R boundary conditions in the open string channel.

In both cases we impose the superconformal invariance at 
the boundary;
\begin{eqnarray}
(L_n-\tilde{L}_{-n})\ket{B}&=&0 \\
(G_r-i\eta \tilde{G}_{-r})\ket{B} &=& 0 \label{G}
\end{eqnarray} 
where $G=G^+ +G^-$. 

Now let us consider the supersymmetric boundary states 
in the $\cN=2$ Liouville theory.
By inspecting the mode expansion of operators of SCA (\ref{SCA-L})
it is easy to see that the A-type and B-type boundary 
conditions require
\begin{itemize}
\item A-type: $\phi$-direction must obey 
the Neumann boundary condition (b.c.), 
and $Y$-direction must be Dirichlet b.c. 
\item B-type: $\phi$-direction must obey the Neumann b.c., 
      and $Y$-direction must be Neumann b.c. 
\end{itemize} 
Note that the Liouville direction $\phi$ must always obey the Neumann
b.c., which means that 
\begin{eqnarray}
  (\alpha^{\phi}_n+\tilde{\alpha}^{\phi}_{-n})\ket{B} &=& 0 ,
      ~~~(n\neq 0)  \\
  \alpha^{\phi}_0\ket{B} &=&\tilde{\alpha}^{\phi}_0\ket{B} =
   \frac{iQ}{2}\ket{B},  \label{bphi}\\
  (\Psi^{\phi}_r + i\eta \tilde{\Psi}^{\phi}_{-r})\ket{B} &=& 0 ,
\hskip4mm \Psi^{\phi}={i \over \sqrt{2}}(\Psi^+ -\Psi^-)
\label{fermion bc}
\end{eqnarray}
where $\dsp i\partial \phi(z) = 
\sum_{n} \frac{\alpha^{\phi}_n}{z^{n+1}}$.
As is  discussed in \cite{Li},
shift $iQ/2$ of Liouville momentum in 
\eqn{bphi} follows from the requirement of conformal invariance 
at the boundary; $(L_n-\tilde{L}_{-n})\ket{B}=0$. 
In fact the commutation relation
\begin{equation}
[L_m, ~ \alpha^{\phi}_n] = -n\alpha^{\phi}_{m+n} -i{Q\over 2}
m(m+1)\delta_{m+n,0},
\end{equation}   
leads to 
\begin{equation}
[L_m-\tilde{L}_{-m},~ \alpha^{\phi}_n + \tilde{\alpha}^{\phi}_{-n}]
= - n (\alpha^{\phi}_{m+n} + \tilde{\alpha}^{\phi}_{-(m+n)}
    -iQ\delta_{m+n,0} ).
\end{equation}
Thus $(\alpha^{\phi}_{0} + 
\tilde{\alpha}^{\phi}_{0})\ket{B} = iQ\ket{B}$. Since
$\phi$, being a non-compact boson, has no winding mode, we have
$\alpha^{\phi}_0=\tilde{\alpha}^{\phi}_0$.
The fact that the supersymmetric boundary states have 
the fixed Liouville momentum $\dsp p_{\phi}=\frac{iQ}{2}$ 
plays an important role in deriving the scaling behavior
of supersymmetric cycles.

It is interesting to consider the Dirichlet condition for the 
Liouville direction 
and this case has been discussed by several authors \cite{Li,RR}. 
However, we only consider
the Neumann condition in this paper.

\subsection{Ishibashi States in Singular $CY_n$ Theory}

Now we explicitly construct the supersymmetric boundary states 
in the singular $CY_n$ theory.
In the following we take the light-cone point of view and 
consider only the transverse directions $\br^{d-2}\times 
(\br_{\phi}\times S^1_Y) \times M_N $ of the theory.
We focus on the A-type boundary condition which corresponds to the
$Dn$-brane wrapped around the middle-dimensional SUSY cycles in $CY_n$
(special Lagrangian submanifold);
\begin{itemize}
 \item All the $\br^{d-2}$-directions $\longrightarrow$ Dirichlet b.c.,
       characterized by the zero-mode momentum 
\begin{equation}
\alpha_0^{\mu}\ket{B}=\tilde{\alpha}_0^{\mu}\ket{B}=p_0^{\mu}\ket{B}
\end{equation}
%%%
 \item $\br_{\phi}$-direction $\longrightarrow$ Neumann b.c., and 
as we have already observed, with the zero-mode momentum fixed as
\begin{equation}
\alpha_0^{\phi}\ket{B}=\tilde{\alpha}_0^{\phi}\ket{B}=\frac{iQ}{2}\ket{B}
\end{equation}
%%%
 \item $S^1_Y$-direction $\longrightarrow$ Dirichlet b.c., characterized 
 by the Kaluza-Klein momentum 
\begin{equation}
\alpha_0^{Y}\ket{B}=\tilde{\alpha}_0^{Y}\ket{B}=p \ket{B} , 
~~ p=\frac{n}{NQ} ~(n\in \bz)
\end{equation}
(The radius of $S^1_Y$ is equal to $NQ$. After imposing GSO projection,
effective radius becomes $Q$ \cite{GKP}).
%%%
 \item $M_N$-sector $\longrightarrow$ A-type boundary condition.
We denote the Ishibashi state \cite{Ishibashi}
in this sector as $\ket{l,m,s}_I$\,
($l=0,1,\ldots, N-2$, $m\in \bz_{2N}$, $s\in \bz_4$, $l+m+s \in 2\bz$)
which is associated to the character $\chi^{l,s}_m(q)$ 
of the representation of the ${\cal N}=2$ minimal model with central charge 
$3(N-2)/N$.
$s$ represents different spin structures of the ${\cal N}=2$ theory.
\end{itemize}

Tensoring all the sectors we can obtain the Ishibashi state 
$\ket{\p_0,s_0;\, l,m,s;\, p}_I$ which diagonalizes
the closed string Hamiltonian $H^{(c)}$
\begin{eqnarray}
&&\hskip-5mm{}_I\bra{\p_0',s_0';\, l',m',s';\, p'} \tilde{q}^{H^{(c)}}
\ket{\p_0,s_0;\, l,m,s;\, p}_I = \frac{\tilde{q}^
{\frac{1}{2}\ps_0^2+\frac{1}{2}p^2}}
{\eta(\tilde{q})^{d}} \chi^{SO(d)}_{s_0}(\tilde{q}) \chi^{l,s}_m(\tilde{q})
\delta(\p_0-\p_0')\delta_{s_0,s_0'}\delta_{l,l'}\cdots, \nonumber \\
&&
\end{eqnarray}
where $\tilde{q}=e^{{-2\pi i\over \tau}}$ and 
$\chi^{SO(d)}_{s_0}(\tilde{q})$ denotes the character of 
$\widehat{SO}(d)_1$ ($s_0=0,1,2,-1$ correspond respectively to 
the basic, spinor, vector, co-spinor representations).
Note that $d-2$ free fermions of the transverse directions $\br^{d-2}$
are combined with the two fermions coming 
from the ${\cal N}=2$ Liouville sector 
and generate the $SO(d)$ current algebra at level 1.

We now impose the GSO conditions
which enforces the integrality of the $U(1)_R$ charge and
determine the spectrum of 
$Y$-momentum $p$ in the Ishibashi state   
$\ket{\p_0,s_0;\, l,m,s;\, p}_I$. 
\begin{itemize}
 \item In all of the cases $d=2,4,6$,
the GSO conditions for the NS sectors is given by
\begin{equation}
-\frac{s_0}{2}- \frac{s}{2} + \frac{m}{N}-Qp \,\in \,2\bz +1 ,~~~(s_0, s= 0,2).
\label{GSO-NS}
\end{equation}
\item For the R sector, the conditions are 
\begin{eqnarray} 
&&d=2  \hskip10mm 
-\frac{s_0}{2}- \frac{s}{2} + \frac{m}{N}-Qp \,\in \,2\bz +1 \label{GSO-R2} \\
&&d=4 \hskip10mm 
-\frac{s_0}{2}- \frac{s}{2} + \frac{m}{N}-Qp \,\in \,2\bz +\frac{1}{2}
\label{GSO-R4} \\
&& d=6 \hskip10mm
-\frac{s_0}{2}- \frac{s}{2} + \frac{m}{N}-Qp \,\in \,2\bz \label{GSO-R6}
\end{eqnarray} 
\end{itemize}
Shift in the right-hand-side of GSO conditions in the R sector is due to the
presence of $U(1)_R$ charge $(d-2)/4$ in Ramond ground state 
of the ${\bf R}^{d-2}$ sector.

\subsection{Cardy States in Singular $CY_n$ Theory}

It is well-known that open string amplitudes factorize when Cardy states 
are used for the boundary states \cite{Cardy}.
Following the standard prescription we construct 
the Cardy state as the "Fourier transform" of an Ishibashi state,
\begin{eqnarray}
&&\hskip-10mm \ket{L, M, S;\, R, X_0,S_0}_C \\
&&=\int d^{d-2}\p_0e^{-i\ps_0\cdot \xs_0} \sum_{l,m,s,s_0,p}
\frac{S_{Ll}}{\sqrt{S_{0l}}}\frac{e^{i\pi \frac{Mm}{N}}}{(2N)^{1/4}}
\frac{e^{-i\pi\frac{Ss}{2}}}{\sqrt{2}}
\frac{e^{-i\pi\frac{S_0s_0}{2}}}{\sqrt{2}}
e^{-ipR}\ket{\p_0,s_0;\, l,m,s; \, p}_I,\nonumber 
\label{Cardy}
\end{eqnarray}
where the sum over $p$ runs over the values dictated by the GSO conditions. 
$S_{L\ell}$ denotes the matrix element of the $S$-transformation of
$\widehat{SU}(2)$ characters 
\begin{equation}
S_{L\ell}=\sqrt{\displaystyle{{2 \over N}}}\sin\left(\displaystyle{{\pi (L+1)
(\ell+1) \over N}}\right).
\end{equation}
$R$ denotes the location around $S^1$ described by the field $Y$. From 
now on we only treat the 
boundary states with the center of mass coordinate $\x_0=0$
and suppress the $\br^{d-2}$ sector.
We then write the Cardy state as $\ket{L,M,S;\, R, S_0}_C$. 

By combining spin structures $S,S_0$ we construct the boundary states 
$\ket{L,M,R;\eta}^{(NS)}_C,\\
\ket{L,M,R;\eta}^{(R)}_C$ as 
\begin{eqnarray}
\ket{L,M,R;+1}^{(NS)}_C &=&  \frac{1}{4}
\sum_{\al,\beta=0,1}
\ket{L,M,2\al;\,R,2\beta}_C, \nonumber \\
\ket{L,M,R;-1}^{(NS)}_C &=& \frac{1}{4}
\sum_{\al,\beta=0,1}\ket{L,M,2\al+1;\,R,2\beta+1}_C, 
\label{CNS} \\
\ket{L,M,R;+1}^{(R)}_C &=& \frac{1}{4}
\sum_{\al,\beta=0,1}(-1)^{\al+\beta}
\ket{L,M,2\al;\,R,2\beta}_C, \nonumber \\
\ket{L,M,R;-1}^{(R)}_C &=& \frac{1}{4}
\sum_{\al,\beta=0,1}(-1)^{\al+\beta}
\ket{L,M,2\al+1;\,R,2\beta+1}_C.
\label{CR}
\end{eqnarray}
One can check that $\ket{L,M,R;\eta}_C^{(NS)}$ in fact contains
only NSNS components of Ishibashi states ($s, s_0=0,2$) and 
similarly $\ket{L,M,R;\eta}_C^{(R)}$ contains the RR components 
($s,s_0=\pm 1$). We also notice the following boundary conditions
in the fermionic sector
\begin{eqnarray}
(G^{\pm}_r -i\eta \tG^{\mp}_{-r}) \ket{L,M,R;\eta}_C^{(*)}&=& 0,  \\
(\Psi^{\pm}_r-i\eta \tilde{\Psi}^{\mp}_{-r}) \ket{L,M,R;\eta}_C^{(*)}&=& 0.
\label{BC fermion}
\end{eqnarray}
We will drop the superscript $(R)$ from  $\ket{L,M,R;\eta}_C^{(R)}$
in the following sections where we will work exclusively in the 
Ramond sector.

%%%%%%%%%%%%%%%%%%%%%%%%%%%%%%%%%%%%%%%%%%%%%%%%%%%%%%%%%%%%%%%%%%%%%%%%%
\section{Boundary State Approach to the Periods}

In this section let us study the periods or the central charges
of the D-branes wrapped around the vanishing cycles. 
We start with simple geometrical computations of
the periods and scaling behaviors of vanishing cycles
when the CY manifold approaches the A-D-E singularities.

\subsection{Geometrical Analysis of Periods}

Let us consider the perturbed Landau-Ginzburg potential
\begin{equation}
F = X^N +\sum_{m=1}^{N-2} g_m X^m +\mu +z_1^2 +\cdots + z_n^2 \equiv 
   P(X;g_m, \mu) +z_1^2 +\cdots + z_n^2.
\label{LGsuper}\end{equation}
$F(X, z_i)=0$ describes the $CY_n$ with an isolated $A_{N-1}$-type
singularity. In the following discussions 
the coupling constants $g_m$ $(\forall m)$ are sometimes 
set to zero for the sake of simplicity.

The holomorphic $n$-form of $X_n$ is defined by
\begin{equation}
\Omega = \frac{dX\wedge dz_1 \wedge \cdots \wedge dz_{n-1}}
{\partial F / \partial z_n}.
\end{equation}
When we assign the $U(1)$-charge 1 to the superpotential $F$,
$n$-form 
$\Omega$ has the $U(1)$-charge $\dsp r_{\Omega}=\frac{Q^2}{2}$.
We also introduce the derivative of the $n$-form in the coupling constant 
$g_m$
\begin{equation}
\Omega_m = \frac{\partial \Omega}{\partial g_m}|_{g_i=0},
\end{equation}
which has the $U(1)$-charge; $\dsp r_{\Omega}+\frac{m}{N}-1$.
Since the variables $z_i,\hskip1mm i=1,\cdots n$ enter quadratically 
in $P(X;g_m,\mu)$ they may be integrated out in the evaluation of periods. 
 Up to a proportionality constant we have
\begin{equation}
\int \Omega=\int P(X;g_m,\mu)^{n-2\over 2}dX=\int P(X;g_m,\mu)^{6-d\over 4}
dX.
\end{equation}

Let us now consider the roots of the equation 
$P(X;g_m,\mu)=0$ and
denote them as $X_a, \hskip1mm a=1,\cdots,N$. We introduce an arbitrary 
path $C_{ab}$ 
in the complex $X$-plane connecting a given pair of roots $X_a$ and $X_b$.
Let us then consider the integral of the $n$-form $\Omega$ 
and compare it with the integral of $|\Omega|$ along the path
$C_{ab}$. Periods are identified as the central charges carried by the 
particles obtained by wrapping the D-brane on vanishing cycles.
On the other hand the integral of the absolute value of the $n$-form
gives the masses of these particles.
We have an obvious inequality
\begin{equation}
\int_{C_{a,b}} |P(X)^{\frac{6-d}{4}} dX| 
\ge \left|  \int_{C_{a,b}} P(X)^{\frac{6-d}{4}} dX \right| \Longleftrightarrow
\mbox{mass}\ge |\mbox{central charge}|.
\label{BPS saturation}
\end{equation}
This inequality is saturated and we obtain BPS states 
iff the phase of $P(X)^{(6-d)/2}dX$ 
is constant 
along the path $C_{ab}$. 
If we suppose, for instance, that the contour $C_{ab}$ is on the real axis 
and $P(X)$ is real and negative along $C_{ab}$.
In this case by taking the real part of the equation $F(X,z_i)=0$
we find a fibration of a sphere $S^{n-1}$ over the interval $C_{ab}$
whose radius vanishes at the end pints $X_a,X_b$. Altogether 
it describes an $n$-dimensional cycle with a minimum volume and thus
a special Lagrangian submanifold of $X_n$ \cite{KLMVW,GVW}. 

In general it is a non-trivial problem to find out if there exists a Lagrangian
submanifold in a given (relative) homology class $[C_{ab}]$. One has to solve
the differential equation
\begin{equation}
 \frac{dX}{dt} = \frac{\al}{P(X)^{\frac{6-d}{4}}},
\label{eq SUSY cycle}
\end{equation} 
and construct an integral curve which passes through both $X_a$ and $X_b$. 
$\al$ in (\ref{eq SUSY cycle}) specifies the phase of the period integral
in question. 
We cannot always expect the existence of 
such an integral curve for an arbitrary deformed polynomial $P(X;g_m,\mu)$
and a pair of roots $X_a$, $X_b$. 
To establish the existence of such a curve is 
an important problem in the study of stable BPS states, and some of its
features have already been discussed in \cite{SV}. 
In our simplest case with the polynomial
$P(X;g_m,\mu)= X^N+\mu$, there exists a unique solution of  
\eqn{eq SUSY cycle} for any choice of $X_a$, $X_b$.
Since each pair $X_a,X_b$ may be identified as a root of $SU(N)$, 
vanishing cycles corresponding to the 
roots of $A_{N-1}$ provide the stable BPS states in our coupling region. 

Let us now parameterize 
pairs of roots of $P(X)=X^N+\mu=0$ as 
\begin{equation}
X_a=(-\mu)^{1/N}e^{i\pi(M+L+1)/N}, \hskip3mm 
X_b=(-\mu)^{1/N}e^{i\pi(M-L-1)/N},
\end{equation} 
where $\dsp L=0,1,\ldots, \left[\frac{N}{2}-1 \right]$,
$M\in \bz_{2N}$, $L+M\in 2\bz+1$.
It turns out that parameters $L,M$ correspond to
the quantum numbers $L,M$ of the ${\cal N}=2$
minimal model. Let us denote the SUSY cycle
corresponding to the pair $(X_a,X_b)$ as $\gamma_{LM}$.

We next evaluate the integral of holomorphic $n$-form
$\Omega$ over a path $C_{LM}$ 
connecting $X_a,X_b$
\begin{equation}
\int_{C_{LM}}(X^N+\mu)^{6-4\over 4}dX
\end{equation} 
Unlike the computation of the mass of a particle, period integral
depends only on the homology class $[C_{LM}]$ and 
is independent of the choice of the contour.  
We may introduce the "Morse function" $W(X)\equiv W(X;\mu)$ by  
\begin{equation}
dW(X) = P(X)^{\frac{6-d}{4}} dX ,
\label{W function}
\end{equation}
and then the period integral is simply expressed as 
\begin{equation}
\int_{C_{LM}} \Omega = \int_{C_{LM}}  P(X)^{\frac{6-d}{4}} dX
  = W(X_b)-W(X_a). 
\label{period ev}
\end{equation}
We note that $W(X)$ is no other than the superpotential for 
the space-time superconformal theory discussed in \cite{GVW}. 
The definition \eqn{W function} leads 
to the following scaling properties of $W(X)$; 
\begin{eqnarray}
W(\al^{1/N}X;\al\mu) &=& \al^{\frac{1}{N}+\frac{6-d}{4}} 
W(X;\mu), \\
W(e^{2\pi i n/N}X;\mu) &=& e^{2\pi i n/N} 
W(X;\mu). 
\end{eqnarray}
We thus obtain 
the following behavior of the periods 
(up to an overall constant independent of $L$, $M$);
\begin{equation}
\int_{C_{LM}} \Omega \approx  \mu^{r_{\Omega}} e^{i\pi \frac{M}{N}}
\sin \left(\pi\frac{L+1}{N}\right) .
\label{period1} 
\end{equation}

Moreover, since the functions $\dsp W_m(X) \equiv 
\frac{\partial W}{\partial g_m}\left|_{g_i=0}\right.$ 
have similar scaling properties,
\begin{eqnarray}
W_m(\al^{1/N}X;\al\mu) &=& \al^{\frac{m+1}{N}+\frac{2-d}{4}} 
W_m(X;\mu), \\
W_m(e^{2\pi i n/N}X;\mu) &=& e^{2\pi i \frac{(m+1)n}{N}} 
W_m(X;\mu) ,
\end{eqnarray}
we also obtain 
\begin{eqnarray}
\int_{C_{LM}} \Omega_m  &\approx & \mu^{r_{\Omega}+\frac{m}{N}-1} 
e^{i\pi \frac{M(m+1)}{N}}
\sin \left(\pi\frac{(L+1)(m+1)}{N}\right), \nonumber \\
&=& \mu^{r_{\Omega}(1-\Delta(g_m))} 
e^{i\pi \frac{M(m+1)}{N}}
\sin \left(\pi\frac{(L+1)(m+1)}{N}\right). \label{period2}
\end{eqnarray}
Here $\dsp \Delta(g_m)=\frac{N-m}{Nr_{\Om}}$
represents the scaling dimension of the coupling constant $g_m$.
In ref. \cite{GVW,SV} the coupling constant $g_m$ is identified as
the VEV of a chiral operator in the space-time superconformal theory.  
For example, in the $d=4$ case $\dsp \Delta(g_m)=2\frac{N-m}{N+2}$,
which agrees with the spectrum of scaling dimensions \cite{EHIY} at the 
Argyres-Douglas points \cite{AD}. 
Since the unitarity bound imposes $\Delta(g_m)> 1$,
$m$ is bounded from above; $\dsp\bar{m}\equiv \frac{N-2}{2}>  m$. 

\subsection{Boundary State Approach to the Periods of Vanishing Cycles}

Let us now turn to the stringy computation of the periods of vanishing
cycles by means of the ${\cal N}=2$ Liouville theory.
We make use of the boundary states constructed in the previous section
and compute their correlation functions by inserting
the cosmological constant operators $S^{\pm}$.
We shall show that the two point function of the boundary state and the 
chiral field depends only on $\mu$ (not on $\bar{\mu}$) 
due to chirality conditions and behave
exactly as in (\ref{period2}).

Following \cite{OOY} we first introduce suitable disk amplitudes which 
correspond to period integrals in
CY manifolds. 
Fix a SUSY cycle $\gamma$, and let $\ket{\gamma}$ be
the corresponding boundary state. We also introduce
a $(c,c)$ chiral primary field $\phi_{\al}$.
We then define the periods of $\gamma$ with respect to $\phi_{\al}$ by
the following two-point function
\begin{equation}
\Pi_{\al}^{\gamma}= \lim_{T\rightarrow +\infty}
\, \btp \phi_{\al}  e^{-TH^{(c)}} \ket{\gamma}_{RR}.
\label{period-da}
\end{equation} 
$\ktpm$, $\btpm$ denote 
the $RR$ vacua obtained from the identity operator 
by the spectral flow.
They possess the hermiticity property
\begin{equation} 
\ktp^{\dagger}=\btm, \hskip3mm  \ktm^{\dagger}=\btp.
\end{equation}
and are expressed explicitly as
\begin{equation}
\begin{array}{l}
 \ktpm=\ket{l=0,m=\pm 1, s=\pm 1}_{M_N}\otimes 
    e^{\mp \frac{i}{2}H_0\pm i\frac{Q}{2}Y_0} \ket{0}_L \\
\hspace{1in}
  \bigotimes \widetilde{\ket{l=0,m=\pm 1, s=\pm 1}}_{M_N}\otimes 
    e^{\mp \frac{i}{2}\tilde{H_0}\pm i\frac{Q}{2} \tilde{Y_0}}
\tilde{\ket{0}}_L, \\
 \btpm=\bra{l=N-2,m=\pm(N-1), s=\pm 1}_{M_N}\otimes 
    \bra{0}_L\, e^{\mp \frac{i}{2}H_0\pm i\frac{Q}{2}Y_0} \\
\hspace{1in}
  \bigotimes \widetilde{\bra{l=N-2,m=\pm (N-1), s=\pm 1}}_{M_N}\otimes 
    \tilde{\bra{0}}_l\,
   e^{\mp \frac{i}{2}\tilde{H_0}\pm i\frac{Q}{2} \tilde{Y_0}}.
\end{array}
\end{equation}
Here the bosonic field $H$ is related to the fermions $\Psi^{\pm}$
in the Liouville sector as $\Psi^{\pm}=e^{\pm iH}$ and $Y_0,H_0$
denote the zero modes of $Y,H$.
These RR ground states are 
annihilated by $G^{\pm}_0$, $\tilde{G}^{\pm}_0$. 

Let us explicitly construct chiral fields $\phi_{\al}$ in the combined 
${\cal N}=2$ Liouville and minimal theories. A general chiral field is
obtained by taking a chiral field of the minimal model $\phi^{M_N}_{m}$
(corresponding to the monomial $X^m$ in the Landau-Ginzburg description) 
and by "gravitationally" dressing it by the Liouville field as
\begin{equation}
\phi_{m,r}=\phi^{M_N}_{m}e^{p_{m,r}(\phi+iY)}, 
\hskip3mm m=0,1,\cdots,N-2,
\hskip2mm r\in{\bf Z}.
\label{primaryLM}\end{equation}
Momentum $p$ of the Liouville and $Y$ field takes values dictated by 
the GSO condition 
\begin{equation}
p=p_{m,r}= \frac{m}{QN}-\frac{r}{Q} 
\label{pmr}
\end{equation}
where $r$ runs over odd integers (in NS sector).
Total $U(1)_R$ charge of $\phi_{m,r}$ is readily computed as
\begin{equation}
\mbox{U(1)-charge}(\phi_{m,r}) =r.
\end{equation}
Fractional charge $\displaystyle{{m\over N}}$ 
from the minimal sector has been canceled 
by the $Y$-momentum of the Liouville sector.
Then $r$ can freely run over (odd) integers.

We can likewise introduce the $(a,a)$ type chiral primary fields
$\phi_{m,r}^*$ and then the states $\phi_{m,r}\ktp$, 
$\phi^*_{m,r}\ktm$ span the complete set of $RR$ vacua.

For each fixed value of $m$ 
it is natural to consider the value $r=r^*$
which gives the minimum $U(1)_R$ charge and identify $\phi_{m,r^*}$
as the "primary operator". Then the other operators 
$\phi_{m,r}\,(r\not =r^*)$  
become its "descendant" fields.  Namely, we adopt $\phi_{m,r^*=1}$ 
as the primary field for each fixed value of $m$.
Quantum number $r$ originates from the momentum around $S^1$ and thus
it represents the KK mode. It may also be identified as the number of 
$D0$ branes attached to the SUSY cycles. These descendant fields appear
quite similar to the gravitational descendants in 2D gravity theory
described by Landau-Ginzburg formulation \cite{TLG}. 
It is curious
to see how far we can push the analogy to 2D gravity.
The appearance of descendants 
has also been noted in $SL(2,{\bf R})/U(1)$ 
formulation in a somewhat different context \cite{LLS}.

We now want to propose that the periods $\Pi^{\gamma}_{\al}$
(\ref{period-da}) with
$\phi_{\al}$ replaced by the primary field $\phi_{m,r^*}$ 
correspond to the periods of 
$\Omega_m$ defined geometrically in (\ref{period2}). 
We will now see how this correspondence works. 

Let us first examine the general properties of the correlation function
\begin{equation}
\Pi^{\gamma}_{\al}=\lim_{T\rightarrow +\infty}
\btp\phi_{m,r}e^{-TH^{(c)}}|L,M,R;\, \eta\rangle_C
\end{equation}
($\al$ stands for $(m,r)$ and $\gamma$ for $(L,M,R)$).
First we should note that the period $\Pi_{\al}^{\gamma}$ 
vanishes in general 
due to Liouville momentum conservation unless the system is perturbed
by Liouville potential terms.
In fact the boundary state has a fixed (imaginary) Liouville momentum 
$p_{\phi}=-Q/2$
while the primary field (\ref{primaryLM}) carries $p_{\phi}=p_{m,r}$ and 
thus the net Liouville
momentum becomes $p_{m,r}-Q/2$ in $\Pi_{\al}^{\gamma}$. 
On the other hand due to the presence of the background charge 
Liouville momentum has to add up to $-Q$ on the sphere in order 
to give a non-vanishing correlation function.
We find that the momentum conservation law 
$p_{m,r}-Q/2=-Q$ has no solution with $m=0,\ldots,N-2,\hskip1mm 
r\in 2{\bf Z}+1$ and thus the period $\Pi_{\al}^{\gamma}$
has to vanish. 

As we have discussed in section 2, there are three different types of
Liouville potentials $S_{nc},S^{\pm}$ which may be used 
for perturbation. It is shown in the appendix that \\
(1) Correlation function $\Pi_{\al}^{\gamma}$ 
remains zero when the non-chiral operator $S_{nc}$ is inserted.
In general $\Pi_{\al}^{\gamma}$ does not depend on the perturbation of the 
K\"{a}hler potential. 
This follows from the chirality properties of the RR vacua 
and also from
the boundary condition satisfied by the boundary state.\\
(2) Correlation function also remains zero under the perturbation by
the anti-chiral operator $S^{-}$. \\
On the other hand, as we shall see in the following, the insertion of 
chiral operator $S^{+}$ into the correlation function gives a non-zero 
result.
Therefore when the
system is perturbed by the cosmological constant operators 
$S^{\pm}$, $\Pi_{\al}^{\gamma}$ depends 
only on $\mu$ and not on $\bar{\mu}$ and thus has the holomorphicity in $\mu$.
In the following we derive the scaling laws of the periods
by inserting the cosmological constant operator $S^{+}$.

Let us now recall some technology of the computation of correlation functions
in Liouville theory. In the standard treatment \cite{Liouville}
one first integrates over the zero-mode of the Liouville field and obtains the
insertion of the Liouville potential term $\sim \Gamma(-n)\mu^n (S^+)^n$
into the correlation function. Here $n$ is the number of insertions of the
cosmological constant and is given by
\begin{equation}
 n=Qp_{m,r}+\frac{Q^2}{2}.
\label{insertion number}
\end{equation}
Note that this agrees with the value given by the Liouville
momentum conservation. We compute correlation functions first by assuming
as if $n$ was an integer and then continue the results 
to the (generically) fractional value (\ref{insertion number}).
This method of analytic continuation is somewhat heuristic, however,
it is known to reproduce the correct results at least in bosonic
Liouville theory. Recently the procedure has been 
made more rigorous by approaches based on the conformal 
bootstrap \cite{L-bootstrap}. 
We assume that the same prescription works also in the case of fermionic
Liouville theory.

Now, the evaluation of the disc amplitude goes as follows: 
As we see in the appendix, insertion of the operator $S^-$ gives a 
vanishing result at $T=\infty$ and we need to consider only the
insertion of the $S^+$ operators. We use the bosonic
representation of fermions $\Psi^{\pm}=e^{\pm iH}$ and the algebra
of vertex operators to obtain
\begin{eqnarray}
&&\hskip-10mm
\dsp \btp \phi_{m,r} (\mu S^+)^n e^{-TH^{(c)}} \ket{L,M,R;\,\eta}_C
=\mu^n\btp \phi^{M_N}_m e^{(p_{m,r}-{n \over Q})
(\phi_0+iY_0+\tilde{\phi}_0+i\tilde{Y}_0)}
  \nonumber \\
&&\hskip-10mm
\times e^{-in(H_0+\tilde{H}_0)}\int\prod_{i=1}^n d^2 z_i \prod_{i<j}|z_i-z_j|^2
\prod_{j=1}^n \tilde{V}^{(+)}(\bar{z}_j)\cdot V^{(+)}(z_j)
e^{-TH^{(c)}}|L,M,R;\,\eta\rangle_C\nonumber \\
&&\label{period eva-a}
\end{eqnarray}
where 
$V^{\pm}(z)=\exp\left(-(\phi^{(\pm)}(z)+iY^{(\pm)}(z))/Q-iH^{(\pm)}(z)
\right)$ and
$\phi^{(+)},Y^{(+)},H^{(+)}$ 
are positive frequency parts of the
fields $\phi,Y,H$. Negative frequency parts of $\phi,Y,H$
have already been annihilated 
by the vacuum $\btp$. We recall that the vacuum $\btp$ carries a $Y$-momentum
$Q/2$ and the boudary state $|L,M;R,\eta\rangle$ has an
(imaginary) $\phi$-momentum
$-Q/2$. Since the net $\phi$ momentum must equal $-Q$ to have a non-zero
correlation function, boundary state $|L,M;R,\eta\rangle$ 
carries an effective 
$\phi$ momentum $Q/2$. We then find the zero mode factors $\phi_0,Y_0$ 
cancell exactly in (\ref{period eva-a})
when the insertion number $n$ takes the value
(\ref{insertion number}).

When $V^{(+)}(z_j)\,(j=1,\ldots,n)$ hit the
boudary state in (\ref{period eva-a}), 
they are converted into negative frequency parts $V^{(-)}$ due to 
the boundary conditions. For the Liouville field we have
\begin{eqnarray}
&&\phi^{(+)}(z) e^{-TH^{(c)}}|L,M,R;\,\eta\rangle=\tilde{\phi}^{(-)}
(z^{-1}e^{-T})
e^{-TH^{(c)}}|L,M,R;\,\eta\rangle,
\end{eqnarray}
and a similar equation holds for $Y,H$ with a $-$ sign in front of 
RHS. We also note that $H$ zero mode obeys a relation
\begin{equation}
e^{-iH_0} |L,M,R;\,\eta\rangle
=i\eta e^{i\tilde{H}_0} |L,M,R;\,\eta\rangle
\end{equation}
in order to reproduce (\ref{BC fermion}).

When we next move these negative frequency operators
to the left so that they get annihilated by the vacuum $\btp$,
we pick up commutator terms with $\tilde{V}^{(+)}(\bar{z}_j)
\, (j=1,\ldots,n)$. Collecting these factors we find 
\begin{eqnarray}
&&\hskip-15mm \Gamma(-n)\btp \phi_{m,r}\left(\mu S^+\right)^n
e^{-TH^{(c)}} \ket{L,M,R;\,\eta}_C  \nonumber \\
&&\hskip-15mm = \dsp  
\frac{c(n)}{2\sqrt{N \sin\left(\frac{\pi(m+1)}{N}\right)}}
\mu^n (i\eta)^n \sin \left(\pi\frac{(L+1)(m+1)}{N}\right) 
 e^{i\pi \frac{M(m+1)}{N}}
\label{period evaluation}
\end{eqnarray}
where we have introduced a notation
\begin{equation}
\hskip-20mm c(n)\equiv \Gamma(-n)e^{-{n(n+1)T\over 2}}
\prod_{i=1}^{n}\int d^2z_i\, \prod_{i<j}|z_i-z_j|^2
\prod_{i,j}{1\over |1-z_i\bar{z}_je^{-T}|^{{2 \over Q^2}+1}}.
\label{Cn}
\end{equation}
Note that the RHS of (\ref{period evaluation}) does not depend on 
$R$ and we will henceforth suppress the label $R$ in the Cardy state.

In the limit of $T\rightarrow +\infty$ we have
\begin{equation}
c(n)=\Gamma(-n)e^{-{n(n+1)T\over 2}}
\prod_{i=1}^{n}\int d^2z_i\, \prod_{i<j}|z_i-z_j|^2 \approx 
e^{-{n(n+1)T\over 2}}A^{n(n+1)\over 2}
\end{equation}
where $A$ is the area of disk (world-sheet). 
The important point here is that the numerical factors 
$c(n),\sqrt{N \sin\left(\frac{\pi(m+1)}{N}\right)}$ 
etc. are independent of the parameters $L$, $M$
characterizing the boundary states. Thus
we may absorb them into the normalization of the chiral
field $\phi_{m,r}$.
After the analytic continuation in $n$ using the selection rule 
\eqn{insertion number}, we obtain 
\begin{equation}
\lim_{T\rightarrow +\infty}\btp \phi_{m,r} e^{-TH^{(c)}} \ket{L,M;\eta}_C
\approx \mu^{\frac{m}{N}+\frac{Q^2}{2}-r} 
\sin \left(\pi\frac{(L+1)(m+1)}{N}\right)
 e^{i\pi \frac{M(m+1)}{N}} .
\label{period-da2}
\end{equation}
Recall that we regard $\phi_{m,r=1}$ as the primary fields.
For these operators we find
\begin{equation}
\lim_{T\rightarrow +\infty}\btp \phi_{m, 1} e^{-TH^{(c)}} \ket{L,M; \eta}_C
\approx \mu^{r_{\Om}(1-\Delta(g_m))} \sin \left(\pi\frac{(L+1)(m+1)}{N}\right)
 e^{i\pi \frac{M(m+1)}{N}} ,
\label{period-da3}
\end{equation}
where $\dsp r_{\Om}\equiv \frac{Q^2}{2}$, 
$\Delta(g_m)= (N-m)/Nr_{\Omega}$.
(\ref{period-da3}) reproduces exactly the geometrical calculation 
\eqn{period2} when we make the identification 
\begin{equation}
\ket{L,M}_C \Longleftrightarrow  \gamma_{L,M}, 
\hskip5mm \phi_{m,1} \Longleftrightarrow \Om_m . 
\label{correspondence}
\end{equation}

We would like to make a few remarks:\\

1. With our convention Ramond ground states are mutually orthogonal
in the sense that
\begin{eqnarray}
&&\btp \phi_{m,r}\phi_{m',r'}\ktp=0, \hskip2mm
\btm \phi_{m,r}^*\phi_{m',r'}^*\ktm=0, \\
&&\btp \phi_{m,r}\phi_{m',r'}^*\ktm\propto \delta_{m,m'}\delta_{r,r'},
\hskip2mm \btm \phi_{m,r}^*
\phi_{m',r'}\ktp\propto \delta_{m,m'}\delta_{r,r'}.
\label{ortho}
\end{eqnarray}
As one may easily check, these properties follow from the conservation laws of 
Liouville and $Y$-momentum and also of 
the fermion number when we evaluate the inner products
by inserting Liouville cosmological terms.\\

2. In view of the result (\ref{correspondence}), 
it is quite natural to suppose that
the marginal perturbation by 
the $(c,c)$ operator
\begin{equation}
\int d^2z \, 
\oint_z dw G^-(w) \oint_{\bar{z}} d\bar{w} \tilde{G}^-(\bar{w}) 
\phi_{m,r^*}(z,\bar{z}),
\end{equation}
corresponds to the deformation of the singularity by a monomial $X^m$ 
\begin{equation}
X^N+\mu \Longrightarrow X^N +g_mX^m +\mu.
\end{equation} 
In order to test this identification
we may compare the norm of the differential form $\Om_m$,
$\dsp \int_{CY_n} |\Om_m|^2$ with 
the Zamolodchikov metric $\btp \phi_{m,1}\phi_{m,1}^* \ktm$
in the moduli space of ${\cal N}=2$ theory. 
Using a similar method of calculation as in the case of periods,
Zamolodchikov metric is evaluated as
\begin{equation}
\begin{array}{l}
\btp \phi_{m,1}\phi_{m,1}^* \ktm  
\approx \btp \phi_{m,1}(\mu S^+)^n e^{-2TH^{(c)}}
 (\bar{\mu}S^-)^n \phi_{m,1}^* \ktm \\
 \hspace{1in} \approx e^{-n(n+1)T}|c(n)|^2 |\mu|^{2r_{\Om}(1-\Delta(g_m))}.
\end{array} 
\label{metric}
\end{equation}
(\ref{metric}) agrees with the geometrical relation
$\dsp \int_{CY_n} |\Om_m|^2 \approx |\mu|^{2r_{\Om}(1-\Delta(g_m))}$.

It is worthwhile to note that the normalizability condition of the operator 
$\phi_{m,r^*}$ at $\phi \rightarrow +\infty$
in the context of Liouville theory
\cite{Seiberg-L} corresponds to the inequality 
$\Delta(g_m)> 1$. This in turn corresponds to the divergence
of the metric
$\dsp \int_{CY_n} |\Om_m|^2 $ in the $\mu \rightarrow 0$ limit.
As discussed in \cite{GVW,SV}, this behavior is equivalent to 
the requirement that $\Om_m$ is 
supported by a normalizable cohomology class localized at the singularity, 
so that the coupling $g_m$ can be realized as the VEV of some dynamical fields 
in space-time conformal theory. 
We summarize 
the normalizability condition of primary fields in various dimensions;
\begin{list}%
 {} %default label
 {} %formatting parameter
 \item $d=6;~$ All the primary fields $\phi_{m,1}$ are normalizable.
 \item $d=4;~$ $\phi_{m,1}$ is normalizable, iff 
        $\dsp 0\leq m < \bar{m}\equiv\frac{N-2}{2}$.
 \item $d=2;~$ All the primary fields $\phi_{m,1}$ are non-normalizable.
\end{list}

3. It turns out that in the conifold case, $d=4$, $N=2$,  
one has $n=0$ ($m=0,r^*=1$) and
the factor $\Gamma(-n)$ coming from the $\phi_0$ integral 
becomes divergent. We then obtain a scaling violation 
\begin{equation}
\mu^{1+\ep} \Gamma(-1-\ep) \sim \frac{\mbox{const}}{\ep}
  +\mu \ln \mu +\cO(\ep).
\end{equation}
Analogy of the conifold theory to 2D gravity coupled to $c=1$ matter has been 
stressed in \cite{GV}.

%%%%%%%%%%%%%%%%%%%%%%%%%%%%%%%%%%%%%%%%%%%%%%%%%%%%%%%%%%%%%%%%%%%%%%%%%%%%%%

\section{Open String Witten Index}

\subsection{Generalities of Open String Witten Index}

Let us next look at intersection numbers among
SUSY cycles in singular CY manifold 
and try to reproduce them using the perturbed Liouville theory.
Intersection numbers among boundary states are computed by
the open string Witten index in world-sheet theory.
Open string Witten index is in general defined as the 
cylinder amplitude \cite{DFiol}
\begin{equation}
I_{\gamma'\gamma} = {}_{RR}\langle\gamma';\eta|\gamma;-\eta\rangle_{RR} ,
\label{Wi}
\end{equation}
where $\gamma$, $\gamma'$ corresponds to the SUSY cycles described 
by the Cardy states.  
Opposite signs for the signatures between in and out states correspond to 
the insertion of $(-1)^F$.
This amplitude can be evaluated by inserting the complete set
of the string states $|n\rangle$ in the RR sector
\begin{equation}
I_{\gamma'\gamma} = \sum_n {}_{RR}\bra{\gamma';\eta} n\rangle 
\langle n\ket{\gamma;-\eta}_{RR} .
\label{Wi2}
\end{equation}
According to the standard argument on supersymmetry index,
one may keep only the RR vacua  
in the intermediate states of this amplitude.
In our formulation, RR-vacua are mutually orthogonal but are 
not normalized. Thus we consider
\begin{equation}
I_{\gamma'\gamma} = \sum_{n\in  RR ~\msc{vacua}} 
\frac{{}_{RR}\bra{\gamma';\eta} n\rangle 
 \langle n \ket{\gamma;-\eta}_{RR}}{\langle n|n\rangle}.
\label{Wi3}
\end{equation}
This gives the starting point of our evaluation of the index.

\subsection{Open String Witten Index for the Singular CY Theory} 

Let us now compute the Witten index 
\begin{equation}
\hskip-5mm I(L',M',R';L,M,R) 
=\sum_{n\in \msc{RR vacua}}\lim_{T\rightarrow \infty}
\displaystyle{{
{}_{C}\bra{L',M',R';+1} e^{-TH^{(c)}}
|n\rangle \times \langle n| e^{-TH^{(c)}}\ket{L,M,R;-1}_C 
\over \langle n| e^{-2TH^{(c)}}|n\rangle }}
\label{Wi4}
\end{equation}
Ordinarily we should be able to compute  
(\ref{Wi4}) for any $T$ and obtain a result which is $T$-independent.
Unfortunately, in Liouville theory we can compute it
only in the limit of $T=+\infty$. 
In the following we evaluate the index at
$T=+\infty$ and assume that 
the result gives the correct topological invariant.

If we use the field 
identification in the minimal sector
\begin{equation}
|\ell,m,s\rangle=|N-2-\ell,m+N,s+2\rangle,
\end{equation}
we can set $s=s_0$ and the RR vacua constructed in the previous section
span the complete set of states to be inserted in (\ref{Wi4}).  
We introduce the following notation 
\begin{eqnarray}
&&|m,r; s'+1\rangle=\phi_{m,r}|0;+\rangle
=e^{p_{m,r}(\phi+iY)}\phi^{M_N}_m|0;+\rangle, \\
&&|m,r; s'=-1\rangle=\phi^*_{m,r}|0;-\rangle
=e^{p_{m,r}(\phi-iY)}{\phi^{M_N}_m}^*|0;-\rangle, \\
&&\langle -m,-r; s'=+1|=\langle 0;+|\phi_{m,r}
=\langle 0;+|e^{p_{m,r}(\phi+iY)}\phi^{M_N}_m, \\
&&\langle -m,-r; s'=-1|=\langle 0;-|\phi_{m,r}^*
=\langle 0;-|e^{p_{m,r}(\phi-iY)}{\phi^{M_N}_m}^*.
\end{eqnarray} 
Here $s'=s=s_0$ and $m$ runs over the values $m=0,\cdots,N-2$.
The numerator of (\ref{Wi4}) is then evaluated as
\begin{eqnarray}
&&\hskip-15mm \dsp{}_{C}\bra{L',M',R';+1} e^{-TH^{(c)}} 
 (\mu_{s'}S^{s'})^{n} \ket{m,r;s'}
 \dsp \approx (\mu_{s'})^{n} e^{-{n(n+1)T\over 2}}  
 \frac{c(n)}{2(2N)^{1/4}} \frac{S_{L',\,m}}{\sqrt{S_{0,\,m}}}\,
 e^{-i\pi\frac{s'(m+1)}{N}M'}, 
\label{index da}\\
&&\hskip-15mm \dsp \bra{-m,-r;-s'}(\mu_{-s'}S^{-s'})^{n}e^{-TH^{(c)}}
\ket{L,M,R;-1}_C  
\dsp \approx (\mu_{-s'})^{n}    e^{-{n(n+1)T\over 2}} 
\frac{c(n)}{2(2N)^{1/4}} \frac{S_{L,\,m}}{\sqrt{S_{0,\,m}}}\,
e^{i\pi\frac{s'(m+1)}{N}M} e^{-i \pi n}. \nonumber \\
&&
\label{index db}
\end{eqnarray}
Here we use the abbreviated notations 
\begin{equation}
\mu_{s'}= \left\{
\begin{array}{ll}
 \mu& (s'=+1) \\
 \bar{\mu}& (s'=-1)
\end{array}
\right. \, ,  ~~~~~
S^{s'}= \left\{
\begin{array}{ll}
 S^+& (s'=+1) \\
 S^-& (s'=-1)
\end{array}
\right. \,.
\end{equation}
The number $n$ of insertions of Liouville potentials is given by  
\begin{equation}
 n=Qp_{m,r}+{Q^2\over 2}
\end{equation} 
as before. 
There is again no $R$, $R'$ dependences in these amplitudes
and we omit these labels from now on.

An additional factor $e^{-i\pi n}$ in 
(\ref{index db}) follows from the fact that
the signature
$\eta$ has opposite signs in the in-coming and out-going
boundary states. Namely, when we use  
the fermionic boundary condition \eqn{BC fermion}, we pick up an additional
$-$ sign for each mode $j=1,\ldots,n$ 
in (\ref{index db}) as compared with (\ref{index da}) and we obtain $(-1)^n$.

We also evaluate the denominator of the formula (\ref{Wi4}) as
\begin{equation}
\bra{-m,-r; -s'} (\mu_{-s'}S^{-s'})^{n}e^{-2TH^{(c)}}
(\mu_{s'}S^{s'})^{n} \ket{m,r;s'} \approx 
e^{-n(n+1)T}c(n)^2 |\mu|^{2n}.
\label{metric-RR} 
\end{equation} 

By combining these calculations the formula \eqn{Wi4}
now reads (up to the overall normalization)
\begin{eqnarray}
I(L',M';L,M) & = & \sum_r \sum_{m=0}^{N-2} \sum_{s'=\pm 1}\,
\frac{S_{L,\,m}S_{L',\,m}}{S_{0,\,m}}\,e^{i\pi \frac{s'(m+1)}{N}(M-M')}
\, e^{-i\pi n}, \nonumber \\
&=& \sum_r \sum_{l=0}^{N-2}\sum_{m=0}^{N-2} \sum_{s'=\pm 1}\,
N^l_{L'\,L} S_{l,\,m}\,e^{i\pi \frac{s'(m+1)}{N}(M-M')}
\, e^{-i\pi n}.\label{n phase}
\end{eqnarray}
$S_{L\,m}$ is the matrix of the 
$S$-transformation of $\widehat{SU}(2)$
and we have used the Verlinde formula for 
$\widehat{SU}(2)_{N-2}$\cite{Verlinde}.
$N^l_{L'\,L}$ denotes the fusion coefficients  
of $\widehat{SU}(2)_{N-2}$.

We find that the factors $c(n),\,e^{-n(n+1)T}$ etc. and 
the powers of $\mu,\,\bar{\mu}$  exactly cancel between the numerator and 
denominator in the formula. This is consistent 
with the fact that we are calculating 
a topological index.

The GSO conditions \eqn{GSO-R2}, \eqn{GSO-R4}, \eqn{GSO-R6}
for $|m,r ;s'\rangle$, $\bra{-m,-r;-s'}$ are now written as 
\begin{eqnarray}
&&d=2,~6; \hskip15mm r  \in 2\bz+1,
\label{GSO-R26-2} \\
&&d=4~; \hskip18mm r \in \left\{\begin{array}{ll}
2\bz \hskip10mm (s'=+1), \\
2\bz +1\hskip3mm (s'=-1).
\end{array}\right.
\label{GSO-R4-2} \end{eqnarray} 
We evaluate the crucial phase factor $e^{-i\pi n}$ in (\ref{n phase})
by using the GSO conditions. We easily find
\begin{eqnarray}
&d=2~;~& e^{-i\pi n} =  e^{-i\pi\frac{(m+1)}{N}},
\\
&d=4~;~& e^{-i\pi n} =  e^{{\pi i \over 2}s'}e^{-i\pi\frac{(m+1)}{N}},
\\
&d=6~;~& e^{-i\pi n} = - e^{-i\pi\frac{(m+1)}{N}}.
\end{eqnarray}
We hence obtain the index for $d=6$,
\begin{eqnarray}
I(L',M';L,M) \hskip-6mm&&= \sum_r \sum_{l=0}^{N-2} \sum_{m=0}^{N-2} \,
N^l_{L'\,L} S_{\ell m}(-e^{i\pi\frac{(m+1)}{N}}) 
(e^{i\pi\frac{(m+1)(M-M')}{N}}+e^{-i\pi\frac{(m+1)(M-M')}{N}}), 
\nonumber\\
&&\hskip-10mm = \sum_r  \sum_{l=0}^{N-2} \sum_{m=0}^{N-2}
N^l_{L'\,L}S_{\ell m}
e^{-i\pi\frac{(m+1)(M-M')}{N}}
(e^{i\pi\frac{(m+1)}{N}}-e^{-i\pi\frac{(m+1)}{N}}), \nonumber \\
&&\hskip-10mm 
\approx \sum_r\sum_{\ell=0}^{N-2}\sum_{\al,\beta=\pm 1}N_{L'L}^{\ell}
(-1)^{\al+\beta\over 2}\delta^{(2N)}(M-M'+\al(\ell+1)+\beta),
\label{Wi-d6} 
\end{eqnarray}
where we have used the relation $\ell+M-M' \equiv L+L'+M-M' \equiv 0 $ 
$(\mod~ 2)$.
$\delta^{(2N)}$ denotes
the Kronecker-delta modulo $2N$.
In the case of $d=2$, calculation of the index 
is completely parallel to the $d=6$ case.

In the case of $d=4$, we instead obtain 
\begin{eqnarray}
I(L',M';L,M) \hskip-6mm &&= 
\sum_r \sum_{l=0}^{N-2} \sum_{m=0}^{N-2}  
N^l_{L'\,L}S_{\ell m} e^{-i\pi\frac{(m+1)}{N}} 
(ie^{i\pi\frac{(m+1)(M-M')}{N}}-ie^{-i\pi\frac{(m+1)(M-M')}{N}}), \nonumber \\
&&\approx \sum_r \sum_{l=0}^{N-2} \sum_{\al,\beta=\pm 1}\,
N^l_{L'\,L}\, (-1)^{\frac{\al}{2}} \,
\delta^{(2N)}(M-M'+\al(l+1)+\beta) \, .
\label{Wi-d4} 
\end{eqnarray}

We have several comments to make:\\

1. The above formulas \eqn{Wi-d6}, \eqn{Wi-d4} involve a decoupled 
sum over $r$, i.e. descendants, which generates an overall 
infinite factor. It is not clear to us if this divergence really exists in the
theory: it is conceivable that these descendants are singular vectors 
in Liouville theory and may not appear as intermediate states in
correlation functions. This is in fact the case in 2D topological 
gravity.
We would like to clarify this issue in a future publication.\\

2. If we ignore  the divergent summation,
our result reproduces the prediction based on the geometrical
calculation under the correspondence \eqn{correspondence}.
It also coincides with the formula recently proposed 
by \cite{Lerche,LLS} using the $SL(2;\br)/U(1)$
Kazama-Suzuki model in the $d=6$ case.  
In these papers, however, the validity of 
a formal analytic continuation in the level
$k$ of $SU(2)_k$ WZW theory to a negative value has been assumed. 
It will be interesting to see if an operator insertion calculation
like ours may be carried out in $SL(2;\br)/U(1)$ theory and confirm 
their result.\\

3. As a consistency check, we notice that 
the matrix $I(L',M';L,M)$ for $d=2,6$
is symmetric while the one for $d=4$ 
is anti-symmetric with respect to $L,\,M$ and $L',\,M'$.
This is consistent with the fact that $I(L',M';L,M)$
is identified as the intersection matrix among 
the even (odd) dimensional cycles for $d=2,6$ ($d=4$).      
      
For example, in the simplest case  $L=L'=0$, we obtain
\begin{eqnarray}
&&\hskip-10mm I_{M,M'}\approx \delta^{(2N)}(M-M'+2)+\delta^{(2N)}(M-M'-2)
        -2\delta^{(2N)}(M-M'), \hskip2mm d=2,6 \\ 
&&\hskip-10mm I_{M,M'} \approx \delta^{(2N)}(M-M'+2)- \delta^{(2N)}(M-M'-2),
\hskip30mm d=4.
\end{eqnarray}       
Especially, in the $d=2,6$ cases we have the extended Cartan matrix 
of the $A_{N-1}$-type \cite{Lerche}.\\

4. Although we have restricted ourselves to the
$A$-type singularity, it is straightforward to extend 
our analysis to the more general A-D-E cases since it is known 
how to construct the Cardy states based on A-D-E type 
modular invariants \cite{BCFT-ADE}.  
In the general A-D-E case our computation of Witten indices 
gives rise to characteristic factors of the form 
$e^{\frac{i\pi m}{h}}-e^{-\frac{i\pi m}{h}}$ ($d=2,6$),
$e^{\frac{i\pi m}{h}}+e^{-\frac{i\pi m}{h}}$ ($d=4$) 
($h$ denotes the Coxeter number of A-D-E algebras) and
we obtain the same results as in \cite{LLS}.

%%%%%%%%%%%%%%%%%%%%%%%%%%%%%%%%%%%%%%%%%%%%%%%%%%%%%%%%%%%%%%%%%%%%%%%%
\section{Relation to the $SL(2;\br)/U(1)$ Approach}

In this section we examine the perturbation of the Liouville theory by the 
non-chiral
operator $S_{nc}$, which corresponds to the screening charge
in the Wakimoto representation of $SL(2,{\bf R})$ current algebra 
\cite{BK}.
As we have pointed out,
this operator does not affect the complex structure of the theory, 
but it modifies its K\"{a}hler structure,
i.e. the metric of the target space. Bosonic part of the action of 
the perturbed theory is given by
\begin{eqnarray}
S_L&=& \frac{1}{8\pi} \int\,d^2z\, \left\{(\partial_a \phi)^2+
(\partial_a Y)^2\right\} -\frac{Q}{8\pi}\int \phi R   \nonumber\\
&&\hspace{1in} +\la\int\,d^2z\, 
(\partial \phi-i\partial Y)(\deebar \phi+i\deebar Y) e^{-Q\phi} .
\end{eqnarray}
Here $\lambda$ is the coupling constant. 
By performing the T-duality transformation along the $Y$-direction, 
we obtain an action
\begin{eqnarray}
S_L^{\msc{dual}}&=& \frac{1}{8\pi} \int\,d^2z\, \left\{(\partial_a \phi)^2+
(\partial_a X)^2\right\} -\frac{Q}{8\pi}\int \phi R   \nonumber\\
&&\hspace{1in} +\la\int\,d^2z\, 
(\partial \phi-i\partial X)(\deebar \phi-i\deebar X) e^{-Q\phi} ,
\label{deformed Liouville action}
\end{eqnarray}
where the dual coordinate is denoted as $X$.
This deformed action \eqn{deformed Liouville action} is 
exactly the one considered in \cite{Eguchi-L} in connection with
the $SL(2,{\bf R})/U(1)$ model of 2D black hole \cite{2DBH}. 
There the coupling contant $\lambda$ 
is interpreted as the black hole mass parameter.
In fact, by changing the coordinates as
\begin{equation}
\left\{
\begin{array}{l}
 \dsp \phi = \frac{2}{Q} \log \cosh r +\phi_0 \\
 \dsp X = \frac{2}{Q}\left(\theta -i \log \tanh r\right)
\end{array}
\right.
\end{equation} 
and tuning  the coupling as  $\la = e^{Q\phi_0}$,
the above action 
is trasformed into
\begin{equation}
S_{\msc{2DBH}} = \frac{k}{4\pi} \int\,d^2z\, \left\{
(\partial_a r)^2+ \tanh^2 r(\partial_a \theta)^2
\right\} -\frac{k}{4\pi}\int R \, \log \cosh r ,
\label{2DBH}
\end{equation}
where we set $\dsp k=\frac{2}{Q^2}$.
This is the exactly the (Euclidean) 2D black-hole
which describes the non-compact cigar geometry. 

Our observation above seems to support the conjecture of \cite{GK} that 
the $SL(2;\br)/U(1)$ model and the $\cN=2$ Liouville theory should be
related by T-duality. Here, however, is a problem:
the counter part of the cosmological
term $S^{\pm}$ is missing in the $SL(2,{\bf R})$ side.  
We do not have an operator in $SL(2,{\bf R})/U(1)$ theory which 
deforms the complex structure of singular CY manifold. 
It seems likely that
the deformed CY manifolds no longer possess $SL(2,{\bf R})$ 
symmetry.

\section{Discussions and Concluding Remarks}

In this paper we have studied the scaling behaviors and intersection
numbers of vanishing cycles in singular CY manifold making use of the
supersymmetric Liouville theory. We have obtained results which are 
in good agreement with geometrical considerations.

It is well-known that in the case of $d=6$ dimensions, the 
$T$-dual of the ALE space is
given by a collection of parallel NS five-branes with its throat region 
being described by the $SU(2)$ WZW model \cite{OV,GHM}. On the other hand 
in the case of 
$d=4$ dimensions we expect the 
$T$-dual of a singular CY threefold to be given by a 
collection of NS five-branes wrapped around a Riemann surface.
Explicit description of such a configuration has not been worked out.  
Instead a configuration of intersecting NS five-branes has been proposed to 
represent the geometry of the conifold \cite{DM}. However, the position of the 
singularity is delocalized in this 
description and intersecting-brane picture
does not seem fit to our framework.

It has recently been pointed
out \cite{PZT} that the geometry of the conifold 
may be approximately 
described by an $\displaystyle{{SU(2)\times SU(2)\over U(1)}}$ WZW model of 
the type proposed in
\cite{GMM}. The sigma model metric of this WZW model is given by 
\begin{equation}
ds^2= k_1(d\theta_1^2+\sin^2 \theta_1 d\phi_1^2)+
k_2(d\theta_2^2+\sin^2 \theta_2 d\phi_2^2)
+k_1(d\psi + \cos \theta_1d\phi_1+\sqrt{k_2\over k_1}\cos\theta_2d\phi_2)^2
\end{equation}
where $k_1$ and $k_2$ are the levels of two $SU(2)$ current algebras.
$\phi_i,\theta_i\,(i=1,2)$ and $\psi$ are the angular variables of  
two and three sphere, $S^2,S^3$.
On the other hand the angular part of conifold metric is given by \cite{CD}
\begin{equation}
ds^2= {1\over 6}(d\theta_1^2+\sin^2 \theta_1 d\phi_1^2)+
{1 \over 6}(d\theta_2^2+\sin^2 \theta_2 d\phi_2^2)
+{1 \over 9}(d\psi + \cos \theta_1d\phi_1+\cos\theta_2d\phi_2)^2.
\end{equation}
Since the target space and the sigma model 
metric of the WZW model should agree only at large $N$ or levels of 
current algebra, we can not unambiguously determine the levels $k_i,i=1,2$
by comparing the above two expressions. However, in our 
construction of modular invariant partition functions \cite{ES}, we
have encountered an extra parafermionic degrees of freedom in $d=4$ theory.
This seems quite consistent with the presence of an extra $SU(2)/U(1)$ factor
in the above construction as compared with the $d=6$ case.
It will be very interesting if it is possible to find a metric for singular 
CY threefolds at large $N$.
 
Very recently a paper hep-th/0011091 by K.Sugiyama and S.Yamaguchi
has appeared which discusses a subject related to this article. 

\section*{Acknowledgement}

T.E. would like to thank H. Ooguri for discussions and Y.S. thanks 
P. Dorey for informative discussion. We also thank SI2000 at Lake Kawaguchi
for providing a stimulating atmosphere.  
Researches of T.E. and
Y.S. are supported in part by the fund for the 
Special Priority Area no.707 "Supersymmetry
and Unified Theory of Elementary Particles" by the Japan Ministry of 
Education. 

%%%%%%%%%%%%%%%%%%%%%%%%%%%%%%%%%%%%%%%%%%%%%%%%%%%%%%%%%%%%%%%%%%%%%%%%%%
\newpage

\noindent
{\bf {\LARGE Appendix}}
\appendix
\section{Holomorphy of Disc Amplitudes}

Let us recall the definition of our disc amplitude 
\begin{equation}
\Pi_{\gamma}^{\al}=\lim_{T\rightarrow +\infty}\btp \phi_{\al} 
 e^{-TH^{(c)}} \ket{\gamma}_{RR} ,
\label{period-app}
\end{equation}
where the $RR$ vacuum $\btp$ is obtained by the spectral flow from the
identity operator in NS sector $\btp = {}_{NS}\bra{0}U$.
$\phi_{\al}$ is a chiral field of $(c,c)$ type.
$\ket{\gamma}_{RR}$ denotes the $RR$-component of the boundary 
state describing the SUSY cycle $\gamma$. 
Note that acting on the vacuum $\btp$, modes of fermionic operators are
shifted as in the topological formulation of the theory 
$U^{-1}G_n^{\pm} U 
= G^{\pm}_{n\mp \frac{1}{2}}$.\\

1. Deformation of the K\"{a}hler moduli\\

The  K\"{a}hler deformations are generated by operators of the form
\begin{equation}
t_a \int d^2z \, 
\oint_z dw G^+(w) \oint_{\bar{z}} d\bar{w} \tilde{G}^-(\bar{w}) 
\hat{\phi}_a(z,\bar{z}) 
+(h.c.).
\label{Kdef}\end{equation} 
Here $\hat{\phi}_a$ is a $(a,c)$ type chiral field. h.c. means the complex 
conjugation and corresponds to the $(c,a)$ field.
In the following we discuss the case of 
$(a,c)$ type operator. Treatment of $(c,a)$ case is exactly similar.

Acting on the vacuum $\btp$ the 1st term of (\ref{Kdef}) is rewritten as
\begin{equation}
t_a \int d^2z \, \left\{ G^+_0,\,[\tilde{G}^-_1 ,
\, \hat{\phi}_a(z,\bar{z})]\right\}.
\end{equation}

We then obtain 
\begin{eqnarray}
&&\btp \{G^+_0,\,[\tG^-_1, \hat{\phi}_a]\} e^{-TH^{(c)}}\ket{\gamma;\eta}
= - \btp \hat{\phi}_a \tG^-_1 G^+_0 e^{-TH^{(c)}}\ket{\gamma;\eta} 
\nonumber \\
&&= -i\eta \btp \hat{\phi}_a \tG^-_1 \tG^-_0  
e^{-TH^{(c)}}\ket{\gamma;\eta}=i\eta \btp \tG^-_0 \tG^-_1 \hat{\phi}_a 
e^{-TH^{(c)}}\ket{\gamma;\eta}=0.
\end{eqnarray}
Thus the period $\Pi_{\gamma}^{\al}$ is independent of
the K\"{a}hler modulus,
\begin{equation}
{\partial \over \partial t_a}\Pi_{\al}^{\gamma}=0.
\end{equation}

In the above we have used the relation 
\begin{equation}
\btp \,G^{+}_n=\btp \,\tG^{+}_n=0  \hskip2mm (n \leq 0),
\hskip3mm \btp \,G^{-}_n=\btp \,\tG^{-}_n=0  \hskip2mm (n \leq 1)
\end{equation}
and the A-type boundary condition
\begin{equation}
(G^{\pm}_0-i\eta \tG^{\mp}_0)\,\ket{\gamma;\eta}=0.
\end{equation}\\

2. Deformation of the complex structure moduli\\

Here we would like to derive the holomorphicity of $\Pi_{\al}^{\gamma}$ 
on the complex structure moduli. 
Let us consider an $(a,a)$ type deformation,
\begin{equation}
\bar{g}_{\al} \int d^2z \, 
\oint_z dw G^+(w) \oint_{\bar{z}} d\bar{w} \tilde{G}^+(\bar{w}) 
\phi^*_{\bar{\al}}(z,\bar{z}) ,
\end{equation} 
where $\phi^*_{\bar{\al}}$ is an $(a,a)$ type anti-chiral field.
Its insertion into the disc amplitude is evaluated as,
\begin{eqnarray}
&&\btp \{G^+_0,\,[\tG^+_0, \phi^*_{\bar{\al}} ]\} e^{-TH^{(c)}}
\ket{\gamma;\eta}
=\btp \phi^*_{\bar{\al}}\tG^+_0 G^+_0 e^{-TH^{(c)}}\ket{\gamma;\eta} 
\nonumber \\
&&= i\eta \btp \phi^*_{\bar{\al}}\tG^+_0 \tG^-_0  
e^{-TH^{(c)}}\ket{\gamma;\eta} 
= i\eta \btp  \phi^*_{\bar{\al}} \{\tG^+_0,\tG^-_0\}   
e^{-TH^{(c)}}\ket{\gamma;\eta} \nonumber \\
&&= i\eta \btp  \phi^*_{\bar{\al}} \left(\tL_0-\frac{\hat{c}}{8}\right)   
e^{-TH^{(c)}}\ket{\gamma;\eta} 
= 0 ~~~\mbox{in the limit $T\rightarrow +\infty$}.
\end{eqnarray}
Hence
\begin{equation}
{\partial\over \partial \bar{\mu}}\Pi_{\al}^{\gamma}=0.
\end{equation}
We can also prove in the same way that the period 
$\bar{\Pi}_{\bar{\al}}^{\gamma}$ defined with respect to 
the vacuum $\btm$ is independent of
the perturbation of the $(c,c)$ type.
Hence we have established that the  holomorphicity of the
period $\Pi^{\gamma}_{\al}$ with respect to the complex structure moduli.

%%%%%%%%%%%%%%%%%%%%%%%%%%%%%%%%%%%%%%%%%%%%%%%%%%%%%%%%%%%%%%%%%%%%%%%%%%

\end{document}